\documentclass[conference]{IEEEtran}
\IEEEoverridecommandlockouts
\usepackage{cite}
\usepackage{amsmath,amssymb,amsfonts}
\usepackage{algorithmic}
\usepackage{graphicx}
\usepackage{textcomp}
\usepackage{balance}
\usepackage{xcolor}
\usepackage{hyperref}
\usepackage[most]{tcolorbox}
\tcbset{textmarker/.style={
        enhanced,
        parbox=false,boxrule=0mm,boxsep=0mm,arc=0mm,
        outer arc=0mm,left=5mm,right=3mm,top=7pt,bottom=7pt,
        toptitle=1mm,bottomtitle=1mm}}
\newtcolorbox{noteBox}{textmarker,
    borderline west={4pt}{0pt}{gray},
    colback=gray!10!white}
\newcommand{\note}[1]{\begin{noteBox} #1 \end{noteBox}}

\def\BibTeX{{\rm B\kern-.05em{\sc i\kern-.025em b}\kern-.08em
    T\kern-.1667em\lower.7ex\hbox{E}\kern-.125emX}}
\begin{document}

\title{From Reality to Virtual Worlds: The Role of Photogrammetry in Game Development}

\author{
    \IEEEauthorblockN{Santiago Berrezueta-Guzman}
    \IEEEauthorblockA{
        \textit{Technical University of Munich} \\
        Heilbronn, Germany \\
        s.berrezueta@tum.de
    }
    \and
    \IEEEauthorblockN{Andrei Koshelev}
    \IEEEauthorblockA{
        \textit{Technical University of Munich} \\
        Heilbronn, Germany \\
        andrei.koshelev@tum.de
    }
    \and
    \IEEEauthorblockN{Stefan Wagner}
    \IEEEauthorblockA{
        \textit{Technical University of Munich} \\
        Heilbronn, Germany \\
        stefan.wagner@tum.de
    }
}
\maketitle

\begin{abstract}

Photogrammetry is transforming digital content creation by enabling the rapid conversion of real-world objects into highly detailed 3D models. This paper evaluates the role of RealityCapture, a GPU-accelerated photogrammetry tool, in game development of Virtual Reality (VR). We assess its efficiency, reconstruction accuracy, and integration with Unreal Engine, comparing its advantages and limitations against traditional modeling workflows. Additionally, we examined user preferences between designed 3D assets and photogrammetry-generated models.
The results revealed that while photogrammetry enhances realism and interactivity, users slightly preferred manually designed models for small, manipulable elements because of the level of detail. However, from a developer perspective, RealityCapture significantly reduces development time while maintaining geometric precision and photorealistic textures. Despite its reliance on high-performance hardware, its automation, scalability, and seamless integration with real-time rendering engines make it a valuable tool for game developers and VR creators. Future improvements in AI-driven optimization and cloud-based processing could enhance accessibility, broadening its applications in gaming, cultural heritage preservation, and simulation.

\end{abstract}


\section{Introduction}\label{I}

In virtual reality (VR) environments, the realism of digital assets directly impacts user immersion and interaction quality, making efficient, high-fidelity modeling a priority \cite{van2019we}. Photogrammetry has become an essential technique in various fields, enabling the conversion of 2D images into highly detailed 3D models \cite{kingsland2020comparative}. Among the leading tools in this domain is \textit{RealityCapture}, a photogrammetry software that leverages GPU acceleration and advanced algorithms to provide rapid, high-accuracy 3D reconstructions \cite{RealityCapture}. RealityCapture has significantly impacted game development, VR, architectural visualization, and cultural heritage preservation projects \cite{kingsland2020comparative, holuvsa2024creating}. Its seamless integration with \textit{Unreal Engine}, a powerful real-time 3D creation tool used in game development, virtual production, architecture, and simulation \cite{UnrealEngine}, has made it a preferred tool for professionals aiming to create realistic and immersive digital environments.

Despite the advancements in photogrammetry, traditional 3D reconstruction workflows face several challenges. Many existing tools suffer from long processing times, limited scalability, and difficulty handling large datasets efficiently \cite{patias2002medical}. Additionally, ensuring high-fidelity textures and maintaining geometric accuracy in real-time applications, such as VR and gaming, remain key hurdles \cite{esmaeili2017workflows}. The complexity of integrating photogrammetry outputs into real-time rendering engines further exacerbates the issue, requiring optimized workflows and seamless interoperability \cite{statham2020use}. 

This paper aims to evaluate the capabilities and applications of photogrammetry in VR game development, focusing on efficiency, reconstruction accuracy, and integration with Unreal Engine. By comparing its advantages and limitations against traditional 3D modeling workflows, we assess how RealityCapture addresses key challenges in photogrammetry and real-time rendering. Specifically, we examine its impact on development time, asset realism, and performance in interactive environments. Additionally, this study investigates user preferences between photogrammetry-generated models and manually designed 3D assets, considering both visual fidelity and interaction quality in VR settings. Through a mixed-methods evaluation involving technical benchmarking and user experience analysis, this research provides practical insights into the trade-offs developers face when selecting asset creation techniques for immersive applications. 

This paper is organized as follows: Section \ref{I} introduces the role of photogrammetry in VR game development, focusing on RealityCapture’s impact on 3D reconstruction. Section \ref{RC} outlines RealityCapture’s development, technical specifications, and features. Section \ref{RW} reviews related work on photogrammetry-based modeling and immersive environments. Section \ref{M} details the study's methodology, including data collection, processing, and evaluation. Section \ref{R} presents the results, while Section \ref{D} discusses key findings, advantages, and limitations. Section \ref{Rec} offers recommendations and best practices for asset creation. Section \ref{C} concludes with a summary and future research directions.

\section{RealityCapture}\label{RC}

RealityCapture is a GPU-accelerated photogrammetry software that converts 2D images and LiDAR scans into high-accuracy 3D models. It was initially developed by Capturing Reality, a company specializing in photogrammetry and computer vision technologies. The software quickly gained recognition for its advanced algorithms and fast processing speed, distinguishing it from competitors. 

In March 2021, Epic Games acquired Capturing Reality, integrating RealityCapture into its ecosystem, particularly for use with Unreal Engine. This acquisition allowed for enhanced development, deeper integration with Epic Games' technology stack, and increased accessibility for game developers. Since its acquisition, RealityCapture has continued to evolve, benefiting from improvements in AI-driven automation, GPU acceleration, and cloud-based reconstruction \cite{aati2020comparative}.

\subsection{Technical Specifications}
\textit{RealityCapture} is designed for high-performance photogrammetry with fast processing speeds while supporting images and laser scans. It achieves this by utilizing \textit{Structure-from-Motion (SfM)}. This technique creates a 3D model of a stationary scene using multiple 2D images  \cite{ozyecsil2017survey}, and \textit{Multi-View Stereo (MVS)}, which refines the model by generating a dense \textit{point cloud} or a triangulated mesh. MVS uses camera positions and sparse 3D points from SfM to build a more complete and accurate representation, allowing RealityCapture to process thousands of images and millions of \textit{point clouds} with high precision \cite{stathopoulou2023survey}.

The software supports multiple input formats, including \textit{JPEG}, \textit{PNG}, \textit{TIFF}, \textit{RAW}, and LiDAR data formats such as \textit{E57} and \textit{LAS}. On the other hand, it exports \textit{OBJ}, \textit{FBX}, \textit{STL}, \textit{PLY}, \textit{XYZ}, and \textit{E57} models, ensuring compatibility with game engines and CAD applications.

\textit{RealityCapture} requires a multi-core CPU, an NVIDIA GPU with CUDA support, and at least 32GB of RAM for optimal performance. It integrates seamlessly with Unreal Engine, leveraging \textit{Nanite},  a virtualized geometry system that renders high-detail assets with minimal performance costs \cite{unreal_nanite} and \textit{Lumen} system, a dynamic global illumination and reflection system designed for next-generation platforms \cite{tan2024mastering}  for advanced real-time rendering. Its Python API and command-line scripting also enable automation and batch processing, enhancing scalability for large projects.

\section{Related Work}\label{RW}

Photogrammetry has gained significant traction in game development, virtual reality (VR), and digital content creation, enabling high-fidelity 3D reconstructions with remarkable accuracy \cite{kasapakis20243d}. Previous research has explored various photogrammetry applications such as RealityCapture, their integration with real-time rendering engines, and their impact on interactive virtual environments \cite{hellmuth2020datasets}.

The study by Yang Pan et al., \cite{pan2022research}, explored photogrammetry and 3D scanning technologies to generate high-fidelity virtual humans for applications in the metaverse. The study discusses integrating Photoscan technology with RealityCapture and Unreal Engine 5 \textit{MetaHuman}, a cloud-based tool for creating high-fidelity, realistic digital humans with advanced animation capabilities, to streamline the creation of realistic digital characters, reducing costs and production time \cite{MetaHuman}. The authors highlighted the significance of photo-based modeling techniques, emphasizing their accessibility, low hardware requirements, and potential for automation. They presented a detailed workflow covering image acquisition, feature matching, mesh optimization, and texture mapping, demonstrating how these methods enhance the realism of virtual avatars. 
Furthermore, the study explored the role of camera arrays in improving the accuracy of human head scans, making digital human production more efficient. The authors also discuss the application of Mesh to MetaHuman technology, which simplifies facial rigging and animation within Unreal Engine, enabling real-time facial capture and motion integration.

The study by Notarangelo et al. explored the integration of photogrammetry and hand gesture recognition (HGR) to create an interactive and immersive virtual walkthrough of \textit{Matera’s Sassi}, a UNESCO World Heritage Site. The study leverages RealityCapture for 3D reconstruction and Unreal Engine for rendering, focusing on enhancing cultural heritage education. The virtual environment allows users to explore historical sites remotely using intuitive hand gestures instead of traditional controllers, improving accessibility and user engagement. 

The authors highlighted the role of deep learning-based HGR systems, which facilitate natural interaction in the digital space through \textit{MediaPipe Hands}, a real-time hand tracking solution developed by Google that detects and tracks hand landmarks using machine learning \cite{zhang2020mediapipe} and a \textit{Feed-Forward Neural Network (FFNN)} trained for real-time gesture recognition. A usability study with 36 participants revealed positive feedback regarding immersion, ease of use, and educational engagement, demonstrating the effectiveness of combining photogrammetry, VR, and HGR for remote learning and cultural heritage preservation. The findings support the growing application of RealityCapture in interactive and educational VR experiences, aligning with advancements in gesture-based navigation and immersive learning environments \cite{notarangelo2023collaborative}.

Maraffi presented a novel pedagogical approach integrating real-time Metahuman animation, photogrammetry, and motion capture as a performing arts process. The framework enables students to create and animate digital doubles using photogrammetry with RealityCapture, Unreal Engine 5, and Autodesk Maya. Unlike traditional keyframing techniques, this approach emphasizes improvisation and embodied acting, leveraging principles from theatre anthropology, Disney’s animation principles, and puppetry. The method fosters creative investment by allowing students to craft digital self-portraits and engage in live scene performance, producing expressive animations that transcend the uncanny valley. This work highlighted the pedagogical potential of integrating performance-based animation workflows in education, offering an alternative to conventional 3D animation pipelines \cite{maraffi2024metahuman}.

These related works suggest that including realistic elements from the real world in VR games can enhance player immersion, engagement, and overall user experience. Real-world-inspired environments, objects, and textures can provide a familiarity that allows players to intuitively navigate virtual spaces and interact with in-game elements. High-fidelity 3D reconstructions can improve the perceived realism of VR applications, reducing the risk of cognitive dissonance and enhancing spatial presence. 

The accurate representation of real-world details, such as natural lighting interactions, material properties, and surface imperfections, further contributes to the believability of virtual environments. This realism is particularly beneficial in applications such as simulation-based training \cite{chen2024task}, serious games \cite{ryan2019photogrammetry, Damianova2025}, architectural visualization \cite{mouratidis2020contemporary, dhanda2019recreating, silva2022reality}, and cultural heritage preservation \cite{pujol2011realism}, where authentic digital twins are essential for conveying accurate spatial and contextual information. As VR technology advances, integrating real-world elements will remain a key factor in bridging the gap between physical and digital experiences, ultimately leading to more engaging, multiplayer, and immersive virtual worlds \cite{newman2022use, berrezueta2025multiplayer}.

\section{Methodology}\label{M}

This study evaluates the trade-offs between automated photogrammetry-based modeling and manual low-poly modeling for VR game development. The focus of this experiment was to compare the efficiency, visual fidelity, and optimization potential of these two approaches by creating digital representations of the same real-world object. The workflow included data acquisition, model reconstruction or manual design, and comparative analysis of the resulting 3D assets.

\subsection{Modeling Approaches}

\begin{figure}[h!]
\centerline{\includegraphics[width=\linewidth]{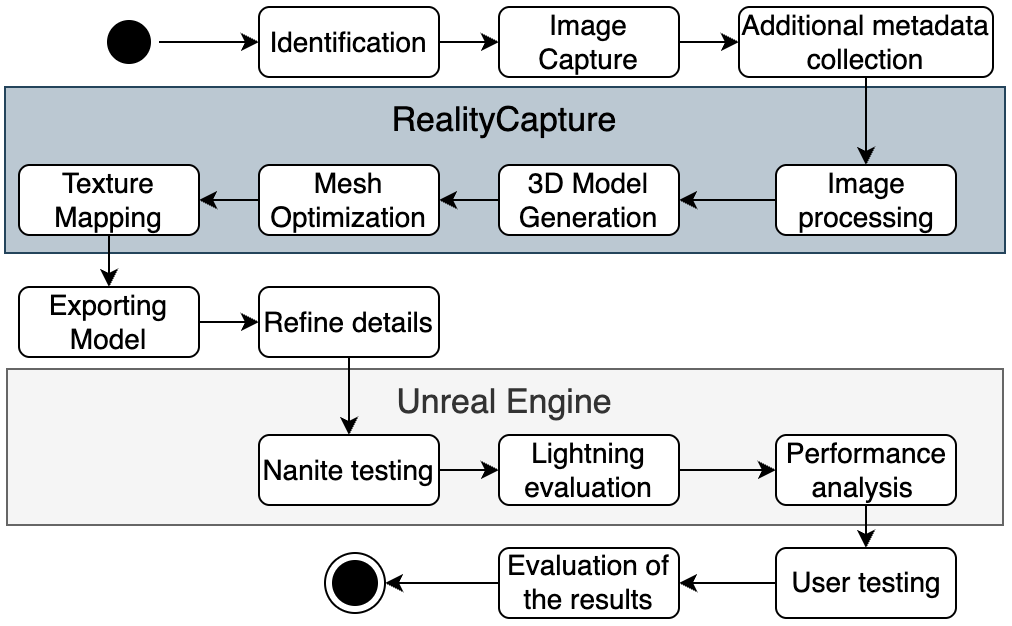}}
\caption{Workflow of image acquisition, RealityCapture-based model creation, the integration in Unreal Engine, and testing process.}
\label{Fig:method}
\end{figure}

Two distinct workflows were used to model the object. The first approach employed photogrammetry using RealityCapture, processing a dataset of 1,550 high-resolution photographs captured from multiple angles to ensure complete surface coverage. RealityCapture used Structure-from-Motion (SfM) and Multi-View Stereo (MVS) algorithms to reconstruct the geometry, producing a high-fidelity mesh with approximately 132,000 vertices and automatically generated textures based on the source images. Figure \ref{Fig:method} illustrates the workflow for photogrammetry using RealityCapture. The second approach involved manual modeling in Blender from scratch. 

\subsection{Comparison Criteria}

The two models were evaluated based on four key aspects:

\begin{itemize}
\item \textbf{Mesh Complexity:} Vertex count and implications for real-time performance.
\item \textbf{Visual Fidelity:} Level of geometric and textural resemblance to the real object.
\item \textbf{Development Time:} Time required for modeling, texturing, and refinement.
\item \textbf{Optimization Requirements:} Effort needed to prepare the model for interactive use in VR environments.
\end{itemize}

\subsection{User Testing and Preference Analysis}

To assess usability and user perception, ten participants interacted with both photogrammetry-generated and pre-designed 3D models in a controlled VR setting. Participants provided feedback on realism, interaction quality, and overall preference. Their preferences were analyzed based on three key metrics:

\textit{- Perceived Realism:} Participants rated the visual fidelity of each model on a Likert scale (1-10).

\textit{- Ease of Interaction:} Users evaluated how naturally and efficiently they could manipulate the objects in VR with the controls. 

\textit{- Immersion and Presence:} Participants indicated whether the model enhanced their sense of realism and engagement in the virtual environment.

Post-experience surveys collected quantitative ratings and qualitative insights. Thematic coding was applied to analyze user feedback, identifying recurring patterns in preferences and interaction experiences. We assessed significant differences between photogrammetry-generated and pre-designed models.

\section{Results}\label{R}

Each method was evaluated based on visual fidelity, mesh complexity, and suitability for real-time applications. 

\subsection{Photogrammetry using RealityCapture}

\begin{figure*}[h!]
\centerline{\includegraphics[width=\linewidth]{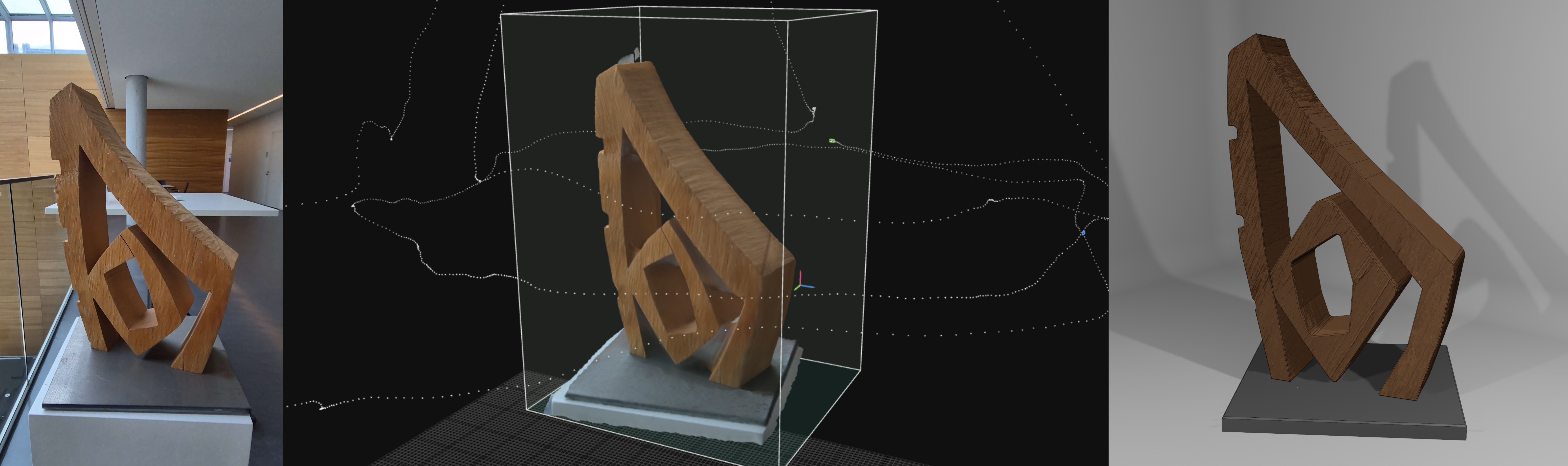}}
\caption{Comparison among (Left) Original photograph of the object. (Center) Reconstructed 3D model using RealityCapture. (Right) The designed model (manually) in Blender.}
\label{Comparison}
\end{figure*}

Photogrammetry, as applied using RealityCapture, enabled the creation of a highly detailed and photorealistic digital replica with minimal manual intervention (see Figure~\ref{Comparison}, center). This approach accurately captured the object’s geometry and surface texture directly from photographs. The white dots visible in the center image represent the camera positions used during the reconstruction process, illustrating the comprehensive coverage required to achieve high visual fidelity.

Regarding reconstruction accuracy and geometric precision, RealityCapture produced high-accuracy 3D reconstructions but minimized the fine details and photorealistic textures. Structure-from-motion (SfM) and Multi-View Stereo (MVS) algorithms helped capture object geometry with minimal distortion. Additionally, the automatic UV mapping and physically-based rendering (PBR) techniques contributed to the enhanced realism of the reconstructed models, making them visually compelling and well-suited for immersive environments. 
However, it is still necessary to refine the model on another platform to increase the realism of the textures and fill parts that RealityCapture could not complete.  

Regarding processing efficiency and performance, the GPU-accelerated reconstruction pipeline significantly reduced processing times compared to traditional photogrammetry workflows.

\subsection{Manual Low-Poly Modeling}

In contrast, the manual low-poly modeling process was completed in under 30 minutes. It yielded a simplified mesh with fewer than 1,000 vertices, making it ideal for real-time use (see Figure~\ref{Comparison}, right). While this approach allowed for fast and lightweight asset generation, it lacked the visual richness of the photogrammetric model. A separate texturing phase was required to approximate the original object's appearance. This highlights a trade-off: manual modeling offers performance efficiency and speed but demands additional effort to approach the realism achieved through photogrammetry. Table \ref{Results} summarizes and compares the results between these methods.

\begin{table}[h]
\caption{Comparison of Photogrammetry and Manual Modeling Approaches}
\label{Results}
\centering
\begin{tabular}{|p{2.1cm}|p{2.6cm}|p{2.9cm}|}
\hline
\textbf{Criterion} & \textbf{Photogrammetry (RealityCapture)} & \textbf{Manual Modeling (Blender)} \\
\hline
Mesh Complexity & \textasciitilde~132,000 vertices & Fewer than 1,000 vertices \\
\hline
Memory Usage (MB)& 1500-1800 & 700-900\\
\hline
Texture Size (MB)& 500-700 & 150-300\\
\hline
Visual Fidelity & High geometric accuracy with textures & Moderate detail, requires manual texturing \\
\hline
Development Time & Several hours (capture and processing) & Approximately 30 minutes (geometry only) \\
\hline
Optimization Requirements & High (mesh simplification needed) & Low (real-time ready) \\
\hline
Rendering& Optimized with Nanite & Less demanding \\
\hline
User Interaction& Require optimization &  Require optimization\\
\hline
\end{tabular}
\end{table}

\subsection{Integration and Performance Evaluation}

\begin{figure}[h!]
\centerline{\includegraphics[width=\linewidth]{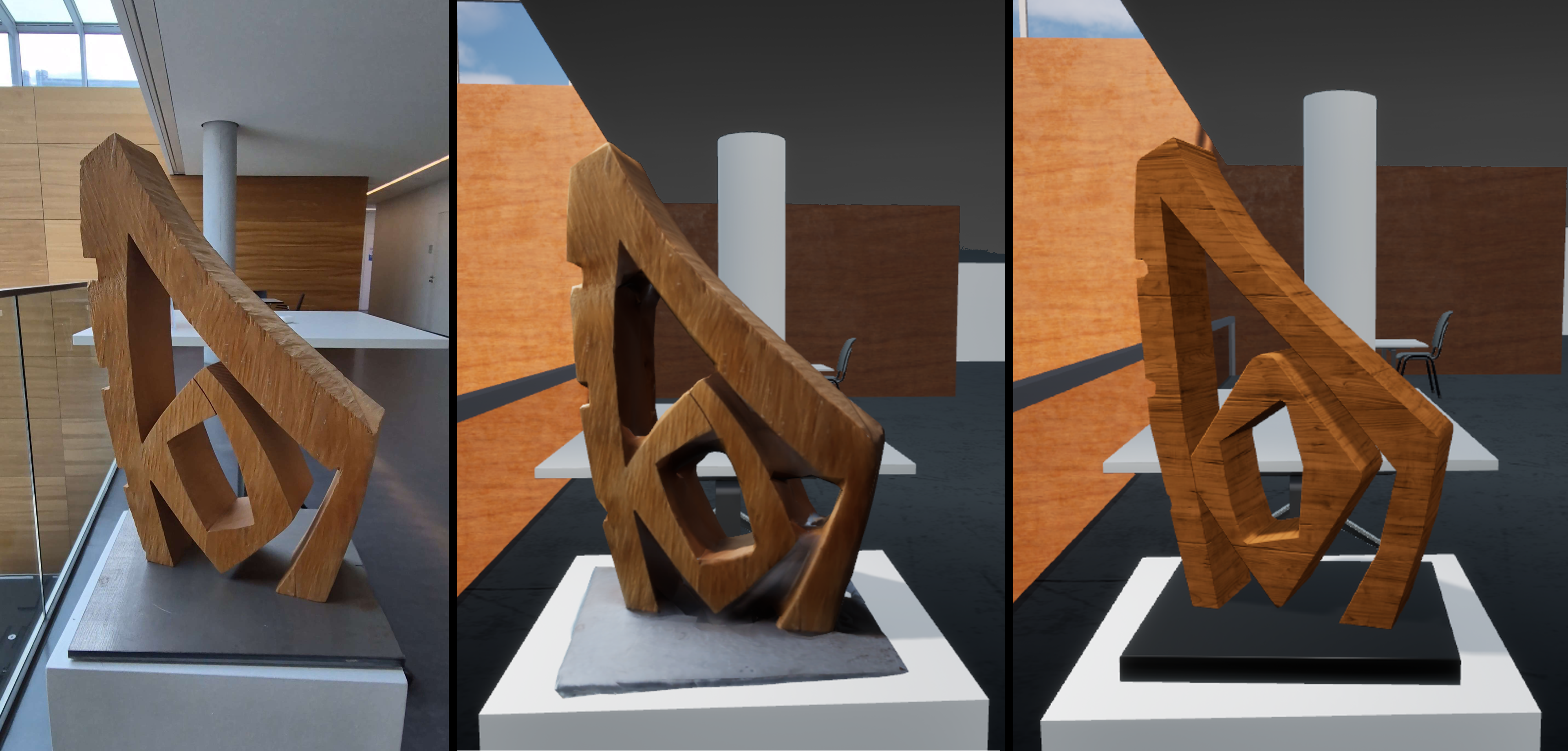}}
\caption{Comparison of different 3D modeling approaches: (Left) Real-world photo of the original sculpture. (Center) RealityCapture-generated 3D model integrated into the virtual environment. (Right) A manually designed 3D model was created in Blender and placed in the same virtual setting.}
\label{ModeledComparison}
\end{figure}

Figure~\ref{ModeledComparison} illustrates a side-by-side comparison of the original sculpture (left), the RealityCapture-generated 3D model (center), and the manually modeled version created in Blender (right), all rendered within the same virtual environment. This setup allowed for a direct assessment of visual fidelity, lighting response, and real-time performance. Both reconstructed models were imported into Unreal Engine, where Nanite rendering was enabled to evaluate the handling of high-resolution assets, particularly for the photogrammetry-based mesh. Lumen was also employed to test global illumination and shadow accuracy under consistent lighting conditions. Performance metrics such as frame rate and memory consumption were recorded, revealing that while the photogrammetric model offered higher visual realism, it required significant optimization to maintain interactive performance compared to the lightweight manually modeled alternative.

\subsection{User Testing and Preference Analysis}

To assess user perception and interaction with the reconstructed models, participants evaluated three key dimensions: perceived realism, ease of interaction, and immersion/presence, each rated on a 10-point Likert scale. The photogrammetry-based model received higher ratings for visual fidelity (M = 8.7, SD = 0.9) compared to the manually modeled version (M = 6.1, SD = 1.2), suggesting a strong preference for the photorealistic textures and geometric accuracy offered by the RealityCapture workflow. 

However, for ease of interaction, the low-poly manually modeled asset scored significantly better (M = 8.4, SD = 0.8) versus the photogrammetric model (M = 5.2, SD = 1.4), primarily due to the reduced mesh complexity and smoother responsiveness in the VR environment. 
Regarding immersion and presence, both models performed similarly, with the photogrammetry model slightly ahead (M = 7.9, SD = 1.0) versus the manual model (M = 7.3, SD = 1.1), though the difference was not statistically significant ($p > 0.05$). These findings highlight a trade-off between realism and interaction performance that developers must consider based on application priorities.

\section{Discussion}\label{D}

\note{\textbf{Finding 1: Trade-off between visual fidelity and performance efficiency.} Balancing the need for photorealism against real-time performance constraints and production efficiency.}

Using 1,550 images in RealityCapture, the photogrammetric method generated a highly detailed and textured model with approximately 132,000 vertices, offering impressive realism with minimal manual input. However, its high mesh complexity posed significant challenges for real-time applications, requiring further optimization. 

In contrast, the manually modeled asset, created in under 30 minutes, featured fewer than 1,000 vertices, making it immediately suitable for interactive environments. Yet, it lacked texture and required an additional phase to achieve comparable visual quality. The intended application should guide the selection of a modeling approach.

\note{\textbf{Finding 2: Limitations and Technical Challenges of RealityCapture.} Despite its many advantages, RealityCapture presented some limitations and challenges.}
High-performance hardware requirements restricted accessibility, making it less practical for developers with lower-end systems. Additionally, complex objects with reflective surfaces, such as metallic or glass materials, posed difficulties in texture reconstruction, often requiring manual corrections. In some cases, further optimization of mesh topology is necessary to improve the usability of photogrammetry-generated models in interactive VR applications.

\note{\textbf{Finding 3: Future Directions and Enhancement Opportunities.} Integrating photogrammetry with procedural generation techniques could streamline asset creation for large-scale virtual environments.}

AI-driven optimization techniques could improve the reconstruction of small interactive objects, enhancing their usability without compromising realism. Cloud-based processing could also mitigate hardware constraints, making the software more accessible to a broader range of developers and enabling more efficient and scalable content production. 

Overall, RealityCapture proved to be a highly effective tool for VR game development, offering significant advantages in reducing model creation time while maintaining high visual fidelity. However, the findings indicate that manually designed models still hold value in scenarios requiring precise interaction mechanics. 

It is important to note that this study involved only ten participants, limiting the findings' generalizability. While the user feedback provided valuable insights into perceived realism, interaction quality, and immersion, the small sample size restricts the statistical power to draw definitive conclusions about broader user preferences. Future research should incorporate larger and more diverse participant groups to capture a wider range of user behaviors, preferences, and interaction patterns. Additionally, repeated trials could help identify trends over time and better assess the impact of photogrammetry-based models on user experience in various VR contexts.

\section{Recommendations}\label{Rec}

The experience gained through this study provides several recommendations for developers and researchers. To maximize the effectiveness of RealityCapture in VR game development and improve photogrammetry-based model creation, we organize our suggestions into the following categories:

\subsection*{Image Acquisition Tips}
\begin{itemize}
    \item Capture high-resolution images with consistent, diffuse lighting to reduce shadows and reflections.
    \item Ensure 60--80\% overlap between images to improve feature matching.
    \item Use a tripod or gimbal to minimize motion blur and noise.
    \item Maintain low ISO and proper shutter speed to enhance image clarity.
\end{itemize}

\subsection*{Workflow Optimization}
\begin{itemize}
    \item Fine-tune reconstruction parameters based on the specific needs of the project.
    \item Use control points and manual markers to improve large-scale scene alignment.
    \item Automate batch tasks using RealityCapture’s Python API to streamline processing.
\end{itemize}

\subsection*{Post-Processing and Refinement}
\begin{itemize}
    \item Simplify high-poly meshes using retopology tools while preserving key surface features.
    \item Bake high-resolution details into normal maps to reduce geometric complexity.
    \item Use tools like Blender or Substance Painter to enhance textures and fix mesh errors.
    \item Test models with Nanite and Lumen in Unreal Engine to validate rendering performance.
    \item For small objects, combine photogrammetry with manual refinement to ensure interactivity.
\end{itemize}

By implementing these recommendations, game developers and researchers can refine photogrammetry-based workflows, improving the realism, efficiency, and accessibility of 3D asset creation in virtual environments.

\section{Conclusions}\label{C}

This study assessed the effectiveness of photogrammetry in VR game development, specifically highlighting RealityCapture and its ability to generate high-quality 3D assets. The results demonstrated that while RealityCapture excels in reconstructing large static objects with high geometric accuracy and photorealistic textures, its performance varies for small interactive elements. Users in this study preferred manually pre-designed models due to their refined detail VR environments. Despite this, RealityCapture remains a valuable tool for game development, significantly reducing production time while maintaining high visual fidelity.

While RealityCapture automates much of the reconstruction process, additional adjustments, such as texture refinement and mesh optimization, are often required to enhance usability. This study also confirmed that seamless integration with real-time rendering engines like Unreal Engine makes RealityCapture an efficient model/asset generation solution in immersive environments.

Future research should explore AI-driven enhancements to photogrammetry workflows for optimizing small object reconstruction and reducing reliance on manual refinements. Cloud-based processing could also improve accessibility, making it more practical for a broader range of developers. Additionally, combining photogrammetry with procedural generation techniques could expand its applications beyond static assets, enabling more interactive and dynamic VR environments.

In conclusion, photogrammetry offers a powerful solution for creating realistic digital models and assets, particularly for large-scale environments in VR game development. While certain limitations remain, advancements in AI, automation, and cloud computing will further enhance its capabilities, making photogrammetry an increasingly integral tool for next-generation virtual content creation.

\section*{Acknowledgment}

This research was financially supported by the TUM Campus Heilbronn Incentive Fund 2024 of the Technical University of Munich, TUM Campus Heilbronn. We gratefully acknowledge their support, which provided the essential resources and opportunities to conduct this study.

\balance
\bibliographystyle{ieeetr}
\bibliography{GEM}

\end{document}